\documentclass[%
 reprint,
 amsmath,amssymb,
 aps,
]{revtex4-2}

\usepackage{amsmath}
\usepackage{quantikz}
\usepackage{hyperref} 
\usepackage{lineno}
\usepackage{graphicx}
\usepackage{slashed}
\usepackage{dcolumn}
\usepackage{bm}

\begin{document}


\preprint{APS/123-QED}

\title{Light front holographic QCD theory in the generalized uncertainty principle framework}
\thanks{A footnote to the article title}%

\author{Fidele J. Twagirayezu}
 \altaffiliation{Department of Physics and Astronomy, University of California, Los Angeles.}
 \email{fjtwagirayezu@physics.ucla.edu}
\affiliation{Department of Physics and Astronomy University of California Los Angeles, Los Angeles, CA, 90095, USA\\
}%


\begin{abstract}
In this article, we develop the framework of light-front holographic QCD in the presence of a minimal length scale by incorporating the Generalized Uncertainty Principle (GUP) into the QCD Lagrangian. From this modified theory, we derive a GUP-corrected light-front holographic QCD (LFH QCD) equation and obtain the corresponding hadronic mass spectrum. Our results show that the hadronic mass spectrum acquires an additional GUP-dependent term that increases the masses. This mass enhancement leads to significantly improved agreement between the theoretical predictions and experimental data.
\end{abstract}

\maketitle


\section{\label{sec:level1}Introduction} 
Quantum Chromodynamics (QCD) provides a robust framework for describing the strong interactions governing quarks and gluons, yet its non-perturbative nature at low energies poses significant challenges for understanding hadron spectroscopy and dynamics. Light-front holographic QCD (LFH QCD) has emerged as a powerful approach to address these challenges by leveraging the AdS/CFT correspondence, mapping strongly coupled QCD dynamics in four-dimensional Minkowski space to a weakly coupled gravitational theory in five-dimensional anti-de Sitter (AdS) space~\cite{Brodsky_2015}. This holographic framework simplifies the treatment of confinement and reproduces key features of the hadronic spectrum, such as linear Regge trajectories.

At short distances, quantum gravitational effects become relevant, potentially altering the fundamental structure of quantum mechanics. The Generalized Uncertainty Principle (GUP) introduces a minimal length scale, often associated with the Planck scale, by modifying the Heisenberg uncertainty principle~\cite{Maggiore:1993kv, Kempf:1996nk, Capozziello:1999wx, Ali:2009zq, Ali_2011,ali2015minimallengthquantumgravity,Bosso:2023aht,Wagner:2023fmb,Twagirayezu:2020chx, Twagirayezu:2021ozk, Twagirayezu:2021kgh,Twagirayezu:2022uow,Twagirayezu:2022ucx}. This minimal length regularizes ultraviolet (UV) divergences and impacts the dynamics of quantum fields, including those in QCD. Incorporating GUP into the QCD Lagrangian offers a pathway to explore quantum gravitational effects on hadronic bound states, particularly in the context of LFH QCD, where the holographic coordinate relates to transverse momentum scales.

In this article, we develop a framework for LFH QCD in the presence of a minimal length scale by integrating GUP into the QCD Lagrangian. We derive a GUP-corrected light-front holographic QCD equation, focusing on the effective Schrödinger-like equation for hadronic bound states. The resulting hadronic mass spectrum includes an additional GUP-dependent term, leading to enhanced masses that improve agreement with experimental data. Our study suggests that light mesons are particularly sensitive to GUP corrections, providing insights into the interplay between quantum gravity and strong interactions.
The organization of this article is as follows:

\section{\label{sec:citeref} Light Front Holographic QCD Theory}

The light-front holographic QCD theory connects Quantum Chromodynamics (QCD) to a holographic framework, specifically using the AdS/CFT correspondence, a strongly coupled QCD dynamics is mapped to a weakly coupled gravitational theory in a higher-dimensional anti-de Sitter (AdS) space. The process requires several steps to bridge the QCD Lagrangian to the light-front holographic QCD (LFH QCD) framework. Below, we outline the key steps to derive the relevant equation, focusing on the effective light-front Schrödinger-like equation for hadronic bound states, which is central to LFH QCD.
\subsection{QCD Lagrangian}
The QCD Lagrangian describes the dynamics of quarks and gluons
\begin{equation}\label{eq:1x}
\begin{aligned}
\mathcal{L}_{\text{QCD}} = \overline{q} (i \slashed{D} - m_q) q - \frac{1}{4} G_{\mu\nu}^a G^{a\mu\nu},
\end{aligned}
\end{equation}
where \( q \) represents quark fields with mass \( m_q \), \( \slashed{D} = \gamma^\mu ( \partial_\mu - i g A_\mu^a t^a ) \) is the covariant derivative, with \( A_\mu^a \) the gluon field, \( g \) the strong coupling, and \( t^a \) the SU(3) color generators,
 \( G_{\mu\nu}^a = \partial_\mu A_\nu^a - \partial_\nu A_\mu^a + g f^{abc} A_\mu^b A_\nu^c \) is the gluon field strength tensor. The indices \( a = 1, \dots, 8 \) run over the gluon color degrees of freedom.

The LFH QCD equation is obtained by using the light-front (LF) coordinates, defined as
\begin{equation}
\begin{aligned}
x^+ = t + z, \quad x^- = t - z, \quad \mathbf{x}_\perp = (x, y),
\end{aligned}
\end{equation}
where \( x^+ \) is the light-front time, and \( x^- \), \( \mathbf{x}_\perp \) are spatial coordinates. The light-front quantization simplifies the vacuum structure and allows us to focus on the dynamics of physical degrees of freedom.
In light-front quantization, the QCD Hamiltonian is derived from the Lagrangian. The energy-momentum tensor \( T^{\mu\nu} \) gives the light-front Hamiltonian
\begin{equation}
\begin{aligned}
P^- = \int d^2 \mathbf{x}_\perp dx^- T^{+-},
\end{aligned}
\end{equation}
where \( P^- \) is the conjugate momentum to \( x^+ \), representing the longitudinal momentum. The goal is to express the dynamics in terms of a light-front Hamiltonian \( H_{\text{LF}} = P^- \), which governs the evolution of the system.

The light-front wave functions (LFWFs) describe the hadronic bound states in terms of their quark and gluon constituents. The eigenvalue problem for a hadron with mass \( M \) is
\begin{equation}
H_{\text{LF}} | \Psi \rangle = M^2 | \Psi \rangle,
\end{equation}
where \( M^2 = P^+ P^- - \mathbf{P}_\perp^2 \), and \( P^+ \) is the total longitudinal momentum. This is a relativistic bound-state equation, but solving it directly in QCD is intractable due to the complexity of the interactions.

Light-front holographic QCD leverages the AdS/CFT correspondence, which posits a duality between a strongly coupled gauge theory (like QCD) in 4D Minkowski space and a weakly coupled gravitational theory in 5D AdS space. The extra dimension in AdS, denoted \( z \), is interpreted as a holographic variable related to the inverse of the light-front transverse momentum scale.

In LFH QCD, the transverse dynamics of hadrons are mapped to a 5D AdS-like theory. The key idea is to approximate the strongly coupled QCD dynamics with a semiclassical theory in AdS space, where the confining potential emerges from the geometry or additional fields.

The AdS metric is
\begin{equation}
\begin{aligned}
ds^2 = \frac{R^2}{z^2} ( \eta_{\mu\nu} dx^\mu dx^\nu - dz^2 ),
\end{aligned}
\end{equation}
where \( R \) is the AdS radius, and \( z \) is the holographic coordinate. The AdS/CFT dictionary relates the boundary (\( z \to 0 \)) to the UV behavior of the gauge theory and the bulk to the IR (confinement) dynamics.

To derive the LFH QCD equation, we consider a 5D effective action in AdS space that captures the dynamics of hadronic modes. For mesons, we typically work with a scalar field \( \Phi(x, z) \) in AdS, representing the meson field. The action is
\begin{equation}
\begin{aligned}
S = \int d^4x \, dz \sqrt{g} \left[ g^{MN} \partial_M \Phi \partial_N \Phi - m_5^2 \Phi^2 - V(\Phi) \right],
\end{aligned}
\end{equation}
where
\( g_{MN} \) is the AdS metric,
 \( m_5^2 \) is the 5D mass, related to the conformal dimension of the boundary operator via \( m_5^2 R^2 = \Delta (\Delta - 4) \), \( V(\Phi) \) is a potential encoding confinement, often introduced via a dilaton or soft-wall model.

For confinement, we use the soft-wall model, where the dilaton profile modifies the action
\begin{equation}\label{eq:7x}
\begin{aligned}
S = \int d^4x \, dz \sqrt{g} e^{-\kappa^2 z^2} \left[ g^{MN} \partial_M \Phi \partial_N \Phi - m_5^2 \Phi^2 \right],
\end{aligned}
\end{equation}
with \( \kappa \) setting the confinement scale (related to \( \Lambda_{\text{QCD}} \)).

\subsection{Equation of Motion and Light-Front Mapping}
Varying the action with respect to \( \Phi \), we obtain the 5D equation of motion. For the soft-wall model, assuming \( \Phi(x, z) = e^{i P \cdot x} \phi(z) \), the equation becomes
\begin{equation}
\begin{aligned}
&\left[ - \partial_z^2 - \hat{\partial}_z + \kappa^4 z^2 - 2 \kappa^2 (2 m_5^2 R^2 - 1) \right] \phi(z) = M^2 \phi(z),\\
&\hat{\partial}_z=\frac{1 - 4 m_5^2 R^2}{z} \partial_z,
\end{aligned}
\end{equation}
where \( M^2 \) is the 4D invariant mass of the meson.

To connect this to light-front dynamics, the holographic coordinate \( z \) is mapped to the light-front transverse separation \( \zeta \), where \( \zeta^2 = x (1 - x) \mathbf{b}_\perp^2 \), with \( x \) the longitudinal momentum fraction and \( \mathbf{b}_\perp \) the transverse separation of constituents. The light-front wave function \( \psi(\zeta) \) is related to \( \phi(z) \) via
\begin{equation}
\begin{aligned}
\phi(z) \sim z^{-3/2} \psi(z), \quad z \leftrightarrow \zeta.
\end{aligned}
\end{equation}

Substituting and simplifying, the AdS equation maps to a light-front Schrödinger-like equation
\begin{equation}
\begin{aligned}
\left[ - \frac{d^2}{d \zeta^2} + V_{\text{LF}}(\zeta) \right] \psi(\zeta) = M^2 \psi(\zeta),
\end{aligned}
\end{equation}
where the effective potential \( V_{\text{LF}}(\zeta) \) is
\begin{equation}
\begin{aligned}
V_{\text{LF}}(\zeta) = \kappa^4 \zeta^2 + \frac{4 L^2 - 1}{4 \zeta^2} + 2 \kappa^2 (J - 1),
\end{aligned}
\end{equation}
with \( L \) the orbital angular momentum, \( J \) the total spin, \( \kappa \) the confinement scale.

This is the light-front holographic QCD equation for the meson spectrum. The potential includes a harmonic oscillator term (\( \kappa^4 \zeta^2 \)) for confinement and a centrifugal-like term (\( \sim 1/\zeta^2 \)) from the AdS geometry.

The eigenvalue \( M^2 \) gives the hadron mass squared. Solving the Schrödinger equation with the potential \( V_{\text{LF}}(\zeta) \), we obtain a Regge-like spectrum
\begin{equation}\label{eq:12x}
\begin{aligned}
M^2 = 4 \kappa^2 (n + L + \frac{J}{2}),
\end{aligned}
\end{equation}
where \( n \) is the radial quantum number. This spectrum reproduces the linear Regge trajectories observed in hadron spectroscopy, a key success of LFH QCD.

The QCD Lagrangian’s non-perturbative dynamics are approximated by the AdS effective action. The confinement potential (\( \kappa^4 \zeta^2 \)) emerges from the soft-wall dilaton, which mimics the IR behavior of QCD. The AdS/CFT duality ensures that the boundary conditions and operator dimensions are consistent with QCD’s UV structure, while the bulk dynamics encode confinement.

To rigorously derive the LFH QCD equation directly from the QCD Lagrangian without AdS/CFT, one would need to solve the full light-front Hamiltonian, including all Fock state contributions, which is computationally infeasible. LFH QCD bypasses this by using the holographic duality to simplify the problem, capturing essential features like confinement and chiral symmetry breaking.

The light-front holographic QCD equation for meson bound states is
\begin{equation}
\begin{aligned}
&\left[ - \frac{d^2}{d \zeta^2} + \kappa^4 \zeta^2 + \frac{4 L^2 - 1}{4 \zeta^2} + 2 \kappa^2 (J-1) \right] \psi(\zeta) \\
&= M^2 \psi(\zeta),
\end{aligned}
\end{equation}
derived by mapping the QCD dynamics to a 5D AdS effective action with a soft-wall dilaton, then projecting onto the light-front coordinates with \( \zeta \sim z \). The potential \( V_{\text{LF}}(\zeta) \) encodes confinement and reproduces the Regge spectrum of hadrons.

\section{Light front Holographic QCD theory with GUP effects}\label{sec:3y}

Incorporating the Generalized Uncertainty Principle (GUP) into the QCD Lagrangian and deriving the light-front holographic QCD (LFH QCD) equation with GUP corrections is a non-trivial task, as GUP modifies the fundamental commutation relations of quantum mechanics, impacting the dynamics at high energies or short distances. Below, we will extend the previous derivation by including GUP corrections in the QCD Lagrangian, then proceed to derive the modified LFH QCD equation, focusing on the effective light-front Schrödinger-like equation for hadronic bound states. The process involves modifying the QCD Lagrangian, adapting the light-front quantization, and adjusting the holographic mapping to account for GUP effects.

The GUP modifies the Heisenberg uncertainty principle to account for quantum gravitational effects, introducing a minimal length scale, often associated with the Planck length. A common form of the GUP-modified commutation relation ~\cite{Bosso_2020}, in the nonrelativistic or semiclassical limit, can be expressed as
\begin{equation}
\begin{aligned}
[x_i, p_j] = i \hbar \left( \delta_{ij} - \beta \hbar^2 p^2 \delta_{ij} - \beta' \hbar p_i p_j \right),
\end{aligned}
\end{equation}
where \( \beta, \beta' \) are GUP parameters with dimensions of inverse mass squared (typically \( \beta \sim 1/M_{\text{Pl}}^2 \), where \( M_{\text{Pl}} \) is the Planck mass), \( p^2 = p_k p^k \).

For simplicity, we adopt the quadratic GUP form (\( \beta' = 0 \))
\begin{equation}
\begin{aligned}
[x_i, p_j] = i \hbar \left( \delta_{ij} - \beta \hbar^2 p^2 \delta_{ij} \right).
\end{aligned}
\end{equation}
This leads to a modified uncertainty relation
\begin{equation}
\begin{aligned}
\Delta x \Delta p \geq \frac{\hbar}{2} \left( 1 + \beta \hbar^2 (\Delta p)^2 \right),
\end{aligned}
\end{equation}
implying a minimal length \( \Delta X_{\text{min}} \sim \sqrt{\beta} \hbar \).

In the context of QCD, GUP corrections are expected to modify the short-distance (UV) behavior of quark and gluon interactions, potentially affecting the dynamics of bound states in the holographic framework.

\subsection{Gauge-covariant GUP-modified QCD Lagrangian }
To incorporate GUP, we need to modify the momentum operators in the Lagrangian of Eq.~\eqref{eq:1x}, as GUP alters the canonical commutation relations. The GUP-modified momentum operator can be approximated as
\begin{equation}
\begin{aligned}
p_i \to \tilde{p}_i = p_i \left( 1 - \beta \hbar^2 p^2 \right),
\end{aligned}
\end{equation}
to first order in \( \beta \), ensuring \( [x_i, \tilde{p}_j] \approx i \hbar ( \delta_{ij} - \beta \hbar^2 p^2 \delta_{ij} ) \). This modification affects the derivative terms in the Lagrangian.
For the quark sector, the Dirac operator \( i \slashed{D} = i \gamma^\mu D_\mu \) becomes
\begin{equation}
\begin{aligned}
D_\mu \to \tilde{D}_\mu  =(1-\beta D_{\rho}D^{\rho})D_{\mu},
\end{aligned}
\end{equation}
where \( D^2 = D_\lambda D^\lambda \). The quark term in the Lagrangian is then
\begin{equation}
\begin{aligned}
\overline{q} (i \gamma^\mu &\tilde{D}_\mu - m_q) q =\\
&\overline{q} \left[ i \gamma^\mu (1-\beta D_{\rho}D^{\rho})D_{\mu} - m_q \right] q.
\end{aligned}
\end{equation}

For the gluon sector, the field strength \( G_{\mu\nu}^a \) involves derivatives of the gluon field. The modified derivative $\tilde{D}_\mu$ introduces higher-order terms
\begin{equation}
\begin{aligned}
\tilde{G}_{\mu\nu}^a = & \partial_\mu A_\nu^a - \partial_\nu A_\mu^a + g f^{abc} A_\mu^b A_\nu^c \\
& - \beta \hbar^2 \left[ (D^\lambda D_\lambda) (\partial_\mu A_\nu^a) - (D^\sigma D_\sigma) (\partial_\nu A_\mu^a) \right].
\end{aligned}
\end{equation}
The gluon term becomes
\begin{equation}
\begin{aligned}
- \frac{1}{4} \text{Tr} \left[ \tilde{G}_{\mu\nu}^{a} \tilde{G}^{a\mu\nu} \right],
\end{aligned}
\end{equation}

The modified QCD Lagrangian becomes
\begin{equation}
\begin{aligned}
\mathcal{L}_{\text{QCD}}^{\text{GUP}} = \bar{q} \left[ i \gamma^\mu \tilde{D}_\mu - m_q \right] q - \frac{1}{4} \text{Tr} \left[ \tilde{G}_{\mu\nu}^{a} \tilde{G}^{a\mu\nu} \right],
\end{aligned}
\end{equation}
These higher-derivative terms complicate the dynamics, introducing non-locality at short distances, consistent with the minimal length scale induced by GUP. Expanding the quark term
\begin{equation}
\begin{aligned}
\tilde{D}_\mu q = D_\mu q - \beta \hbar^2 (D^\lambda D_\lambda) D_\mu q,
\end{aligned}
\end{equation}

introduces higher-order gauge interactions, preserving gauge invariance since \( D_\mu \) is covariant. The gluon term involves complex higher-derivative interactions due to the commutator, ensuring SU(3) covariance.

\subsection{Light-Front Quantization with GUP}
In light-front coordinates (\( x^+, x^-, \mathbf{x}_\perp \)), the Lagrangian is quantized using the light-front Hamiltonian \( P^- \), derived from the energy-momentum tensor \( T^{+-} \). The GUP modification affects the momentum operators in the light-front framework. The light-front momentum components are
\begin{equation}
\begin{aligned}
p^+ = p^0 + p^3, \quad p^- = p^0 - p^3, \quad \mathbf{p}_\perp = (p^1, p^2).
\end{aligned}
\end{equation}
The GUP-modified commutation relations in light-front coordinates are more complex due to the non-standard metric, but we approximate the transverse momentum operators as
\begin{equation}
\begin{aligned}
p_i^\perp \to \tilde{p}_i^\perp = p_i^\perp \left( 1 - \beta \hbar^2 (\mathbf{p}_\perp^2 + (p^-)^2) \right),
\end{aligned}
\end{equation}
assuming the dominant GUP effects are in the transverse plane, where holographic variables are defined.

The light-front Hamiltonian \( H_{\text{LF}} = P^- \) is modified by the GUP terms in the Lagrangian. The quark kinetic term, for example, includes
\begin{equation}
\begin{aligned}
\overline{q} i \gamma^+ \partial_- (1 - \beta \hbar^2 \partial^2) q,
\end{aligned}
\end{equation}
where \( \partial_- = \partial / \partial x^- \). The gluon self-interaction terms are similarly modified, leading to a corrected Hamiltonian
\begin{equation}
\begin{aligned}
H_{\text{LF}}^{\text{GUP}} = H_{\text{LF}} + \beta \hbar^2 H_{\text{c}},
\end{aligned}
\end{equation}
where \( H_{\text{c}} \) includes higher-derivative interactions.

The eigenvalue problem for hadronic bound states remains
\begin{equation}
\begin{aligned}
H_{\text{LF}}^{\text{GUP}} | \Psi \rangle = M^2 | \Psi \rangle,
\end{aligned}
\end{equation}
but the GUP corrections modify the effective potential and kinetic terms in the light-front wave function (LFWF) dynamics.
\subsection{AdS/CFT Correspondence and Holographic Mapping with GUP}
In LFH QCD, the AdS/CFT correspondence maps QCD dynamics to a 5D AdS space, with the holographic coordinate \( z \) related to the light-front transverse separation \( \zeta \). The GUP introduces a minimal length scale, which can be incorporated into the AdS framework by modifying the AdS metric or the effective action to account for UV modifications.

We use the soft-wall model, where the action for a scalar field \( \Phi(x, z) \) (representing mesons) is given by Eq.~\eqref{eq:7x}. 
The GUP correction can be modeled by modifying the kinetic term to include higher-derivative terms, reflecting the non-locality
\begin{equation}\label{eq:29x}
\begin{aligned}
\partial_M \to \tilde{\partial}_M = \partial_M \left( 1 - \beta \hbar^2 \square_5 \right),
\end{aligned}
\end{equation}
where \( \square_5 = g^{MN} \partial_M \partial_N \) is the 5D d’Alembertian. The modified action becomes
\begin{equation}\label{eq:30x}
\begin{aligned}
S_{\text{GUP}} = \int d^4x \, dz \sqrt{g} e^{-\kappa^2 z^2}
\left[ g^{MN} \tilde{\partial}_M \Phi \tilde{\partial}_N \Phi - m_5^2 \Phi^2 \right].
\end{aligned}
\end{equation}

Alternatively, GUP effects can be incorporated by modifying the AdS metric to include a UV cutoff at \( z \sim \sqrt{\beta} \), or by adding a potential term \( V_{\text{GUP}}(z) \) that suppresses dynamics at small \( z \). For simplicity, we proceed with the modified kinetic term.

\subsection{Equation of Motion with GUP}
Varying the modified action, the equation of motion for \( \Phi(x, z) = e^{i P \cdot x} \phi(z) \) is
\begin{equation}
\begin{aligned}
&\biggl[ - \partial_z^2 \left( 1 - \beta \hbar^2 \partial_z^2 \right)^2 - \frac{1 - 4 m_5^2 R^2}{z} \partial_z +\kappa^4 z^2 \\
&- 2 \kappa^2 (2 m_5^2 R^2 - 1) \biggr] \phi(z) = M^2 \phi(z).
\end{aligned}
\end{equation}
The GUP term \( (1 - \beta \hbar^2 \partial_z^2)^2 \) introduces a fourth-order derivative, reflecting the non-local nature of GUP. To map this to the light-front framework, we relate \( \phi(z) \sim z^{-3/2} \psi(\zeta) \), with \( z \leftrightarrow \zeta \). The transverse kinetic term in the light-front Hamiltonian is modified as
\begin{equation}\label{eq:32x}
\begin{aligned}
-\frac{d^2}{d \zeta^2} \to -\frac{d^2}{d \zeta^2} \left( 1 - \beta \hbar^2 \frac{d^2}{d \zeta^2} \right)^2.
\end{aligned}
\end{equation}
The resulting light-front Schrödinger-like equation is
\begin{equation}
\begin{aligned}
\left[ - \frac{d^2}{d \zeta^2} \left( 1 - \beta \hbar^2 \frac{d^2}{d \zeta^2} \right)^2 + V_{\text{LF}}(\zeta) \right] \psi(\zeta) = M^2 \psi(\zeta),
\end{aligned}
\end{equation}
where the potential \( V_{\text{LF}}(\zeta) \) retains its form from the non-GUP case
\begin{equation}
\begin{aligned}
V_{\text{LF}}(\zeta) = \kappa^4 \zeta^2 + \frac{4 L^2 - 1}{4 \zeta^2} + 2 \kappa^2 (J - 1).
\end{aligned}
\end{equation}
Now, we are required to find the solution to the GUP-modified Equation. The GUP term \( \beta \hbar^2 \frac{d^4}{d \zeta^4} \) (and higher-order terms) complicates the solution. To first order in \( \beta \), we approximate the kinetic term as
\begin{equation}
\begin{aligned}
-\frac{d^2}{d \zeta^2} \left( 1 - 2 \beta \hbar^2 \frac{d^2}{d \zeta^2} \right).
\end{aligned}
\end{equation}
The equation becomes
\begin{equation}
\begin{aligned}
&\biggl[ - \frac{d^2}{d \zeta^2} + 2 \beta \hbar^2 \frac{d^4}{d \zeta^4} + \kappa^4 \zeta^2 \\
&+ \frac{4 L^2 - 1}{4 \zeta^2} + 2 \kappa^2 (J - 1) \biggr] \psi(\zeta) = M^2 \psi(\zeta).
\end{aligned}
\end{equation}
The fourth-derivative term modifies the UV behavior, effectively introducing a cutoff at small \( \zeta \), consistent with the minimal length scale.

Solving this equation analytically is challenging due to the higher-order derivatives. Numerically or perturbatively, the GUP correction shifts the eigenvalues \( M^2 \). For small \( \beta \), the mass spectrum is modified as
\begin{equation}\label{eq:36x}
\begin{aligned}
M^2 \approx 4 \kappa^2 \left( n + L + \frac{J}{2} \right) + \delta M^2,
\end{aligned}
\end{equation}
where the quantity \( \delta M^2 \) arises from the GUP term, typically increasing the mass slightly due to the suppression of short-distance interactions.
Using first-order perturbation theory,
the corrected light front wavefunction to first order in \( \beta \) is
\begin{equation}
\begin{aligned}
\psi_n = \psi_n^{(0)} + \beta \sum_{m \ne n} c_{nm} \psi_m^{(0)} + \mathcal{O}(\beta^2).
\end{aligned}
\end{equation}
The correction to the eigenvalue is
\begin{equation}\label{eq:38x}
\begin{aligned}
\delta M^2 = \langle \psi_n | \hat{H}' | \psi_n \rangle = 2\beta \hbar^2 \langle \psi_n | \frac{d^4}{d \zeta^4} | \psi_n \rangle
\end{aligned}
\end{equation}
To estimate this matrix element, note that for harmonic oscillator-like wavefunctions \( \psi_n(\zeta) \), we know
\begin{equation}
\begin{aligned}
& \langle \psi_n | \zeta^2 | \psi_n \rangle \sim (n + L + 1/2)/\kappa^2, \\
&\langle \psi_n | -\frac{d^2}{d\zeta^2} | \psi_n \rangle \sim \kappa^2(n + L + 1/2). 
\end{aligned}  
\end{equation}
So the fourth derivative roughly scales as
\begin{equation}
\begin{aligned}
\left\langle \frac{d^4}{d\zeta^4} \right\rangle \sim \kappa^4 (n + L + 1/2)^2
\end{aligned}
\end{equation}
Therefore Eq.~\eqref{eq:38x} becomes
\begin{equation}
\begin{aligned}
\delta M_n^2 \sim \beta \hbar^2 \kappa^4 (n + L + 1/2)^2
\end{aligned}
\end{equation}
If we now include spin (e.g., via \( J = L + S \)), the combination \( (n + L + S/2) \) naturally appears, and the expression becomes
\begin{equation}
\begin{aligned}
\delta M^2 \sim \beta \hbar^2 \kappa^4 \left(n + L + \frac{S}{2}\right)^2
\end{aligned}
\end{equation}

The GUP correction introduces a minimal length scale, which regularizes the UV divergences in QCD and modifies the short-distance behavior of the light-front wave functions. In the holographic context, this corresponds to a cutoff in the AdS space at small \( z \), consistent with the expectation that quantum gravity effects become relevant near the Planck scale. The confinement potential \( \kappa^4 \zeta^2 \) remains dominant in the IR, so the Regge-like spectrum is preserved, with small GUP-induced shifts.

\section{Consistency of GUP Implementation with AdS/QCD Duality}

The light-front holographic QCD (LFHQCD) framework relies on a semiclassical correspondence between a 5D gravitational theory in anti-de Sitter (AdS) space and strongly coupled QCD in physical spacetime. A key aspect of this duality is that UV behavior in QCD (short-distance physics) is captured by the boundary behavior (small $z$) of the AdS fields, while IR dynamics (confinement) emerge from the bulk structure or the dilaton field.

The inclusion of the Generalized Uncertainty Principle (GUP) into the LFHQCD framework modifies the fundamental commutation relations and introduces a minimal length scale, typically associated with short-distance (UV) regularization. It is therefore essential to ensure that these modifications do not disrupt the basic UV–IR mapping established by the AdS/QCD correspondence.

\subsection{Mapping GUP Corrections to AdS Space}

In our implementation, GUP corrections are introduced through higher-derivative terms in both the 4D QCD Lagrangian and its 5D holographic dual. In the AdS effective action, this is reflected by the replacement of the derivative operator as indicated by Eq.~\eqref{eq:29x}. This leads to a modified kinetic term in the AdS action in Eq.~\eqref{eq:30x}. The higher-derivative correction $\Box_5$ becomes significant at small $z$, mimicking a UV cutoff in the AdS geometry. Importantly, the GUP-deformed action preserves the general structure of the AdS background and the soft-wall dilaton profile $\phi(z) = \kappa^2 z^2$, which ensures confinement and the emergence of linear Regge trajectories.
\subsection{Preservation of UV–IR Correspondence}
The small-$z$ behavior of the scalar field $\Phi(z)$ near the AdS boundary still satisfies the expected asymptotic form,\(
\Phi(z) \sim z^\Delta, \quad \text{as } z \to 0,\)
with $\Delta$ the conformal dimension determined by the AdS mass via $m_5^2 R^2 = \Delta(\Delta - 4)$. The GUP corrections introduce higher-order derivatives in $z$, but do not alter the leading behavior of $\Phi(z)$ unless $\beta$ is large enough to invalidate the perturbative expansion. In our model, the fit parameter $\beta $ is chosen to be sufficiently small to preserve the asymptotic freedom regime in QCD, thus maintaining the correct UV behavior at the boundary.
\subsection{Consistency of Potential Structure}

In the light-front Schrödinger equation, the GUP modifies only the kinetic term as indicated in Eq.~\eqref{eq:32x} while the effective potential $V_{\text{LF}}(\zeta)$, derived from the AdS geometry and dilaton profile, remains unchanged. Therefore, the duality-preserving potential structure is left intact, ensuring that confinement and the IR structure of the theory are preserved.

\subsection{Gauge/Gravity Correspondence Compatibility}

The overall GUP implementation modifies the dynamics only in the UV regime, introducing a minimal length and suppressing small-$\zeta$ (or small-$z$) contributions. Since the AdS/QCD duality is a low-energy effective correspondence — not a complete quantum gravity embedding — such effective UV modifications are compatible with the duality as long as the large-$z$ (IR) structure, potential form, and operator scaling dimensions are unmodified at leading order.
Thus, we conclude that the incorporation of GUP through a higher-derivative kinetic correction preserves the geometry and scale separation of the AdS/QCD correspondence, retains the correct operator scaling near the UV boundary, and maintains the form of the confinement-inducing potential in the IR.

Thus, the GUP deformation, implemented as a UV-effective correction in both the 4D and 5D formulations, preserves the essential structure and consistency of the AdS/QCD duality. It enhances the UV behavior without breaking the confinement physics encoded in the AdS geometry or soft-wall dilaton, allowing a consistent holographic interpretation of minimal-length effects in hadronic physics.

\section{Phenomenology}
\begin{figure}[htb]
    \centering    \includegraphics[scale=0.425]{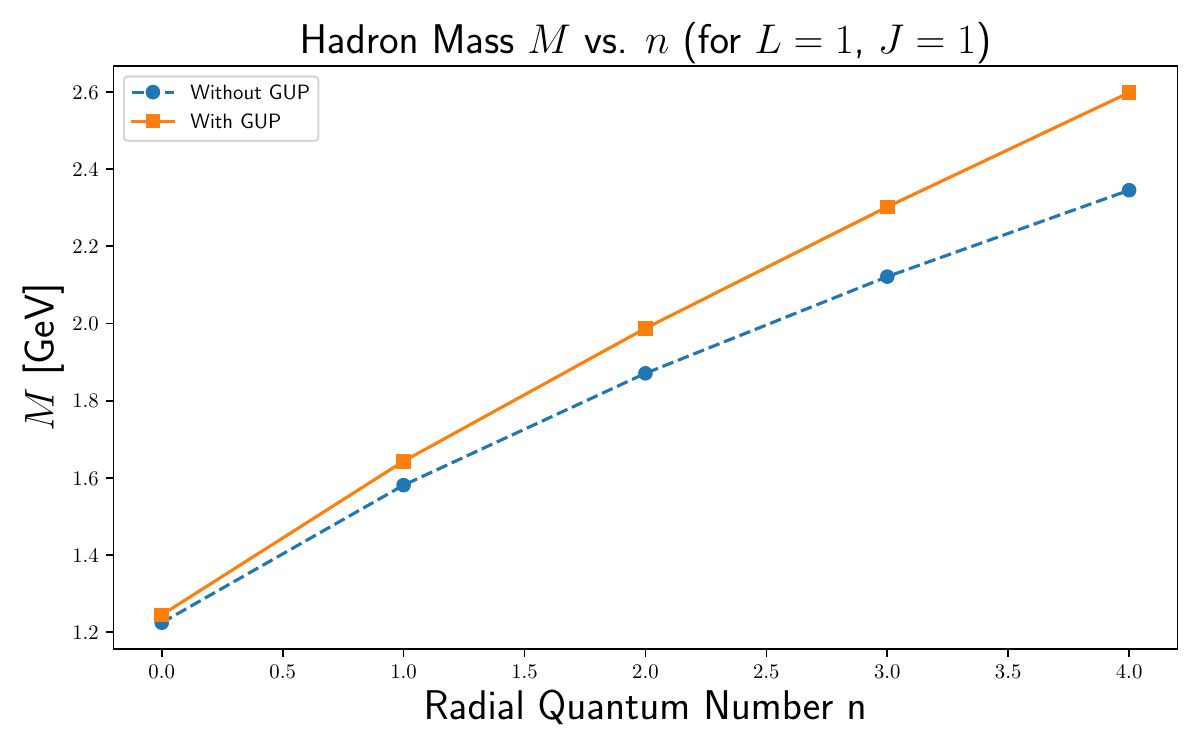}
    \caption{The hadron mass spectrum with and without GUP corrections is plotted as function of the radial quantum number $n$. The value of $\kappa$ is set to 0.5, and $\beta$ is assigned a value of 0.9 GeV$^{_2}$, chosen arbitrarily for illustration purposes.}
    \label{fig:1}
\end{figure}

In this section, we present phenomenological results comparing scenarios with and without GUP corrections. Here, we mainly focus on the mass spectrum given by Eqs.~\eqref{eq:12x} and ~\eqref{eq:36x}. First, we show the effect of the GUP corrections on the mass spectrum by plotting $M$ versus $n$ using Eqs.~\eqref{eq:12x} and \eqref{eq:36x}. Figure~\eqref{fig:1} shows the plot of the hadron mass \( M \) versus radial quantum number \( n \) for fixed \( L = 1 \) and \( J = 1 \), comparing results with and without GUP corrections using arbitrary $\beta$ value. The GUP-corrected masses increase more rapidly with \( n \), showing the expected shift from the higher-derivative (UV-suppressing) terms. 

\begin{table}[h!]
\centering
\begin{tabular}{|c|cccccc|}
\hline
\textbf{Meson} & $n = 0$ & $n = 1$ & $n = 2$ & $n = 3$ & $n = 4$ & $n = 5$ \\
\hline
$\pi$ (Pion) & 0.135 & 1.3 & 1.8 & 2.1 & 2.4 & 2.6 \\
$\rho$ (Rho) & 0.775 & 1.45 & 1.7 & 2.0 & 2.3 & 2.5 \\
\hline
\end{tabular}
\caption{Experimental meson masses (in GeV) used for $\pi$ and $\rho$ states up to $n=5$, collected from the Particle Data Group(PDG)~\cite{PDG2022}.}
\label{tab:meson_masses}
\end{table}

\begin{figure}[htb]
    \centering    \includegraphics[scale=0.325]{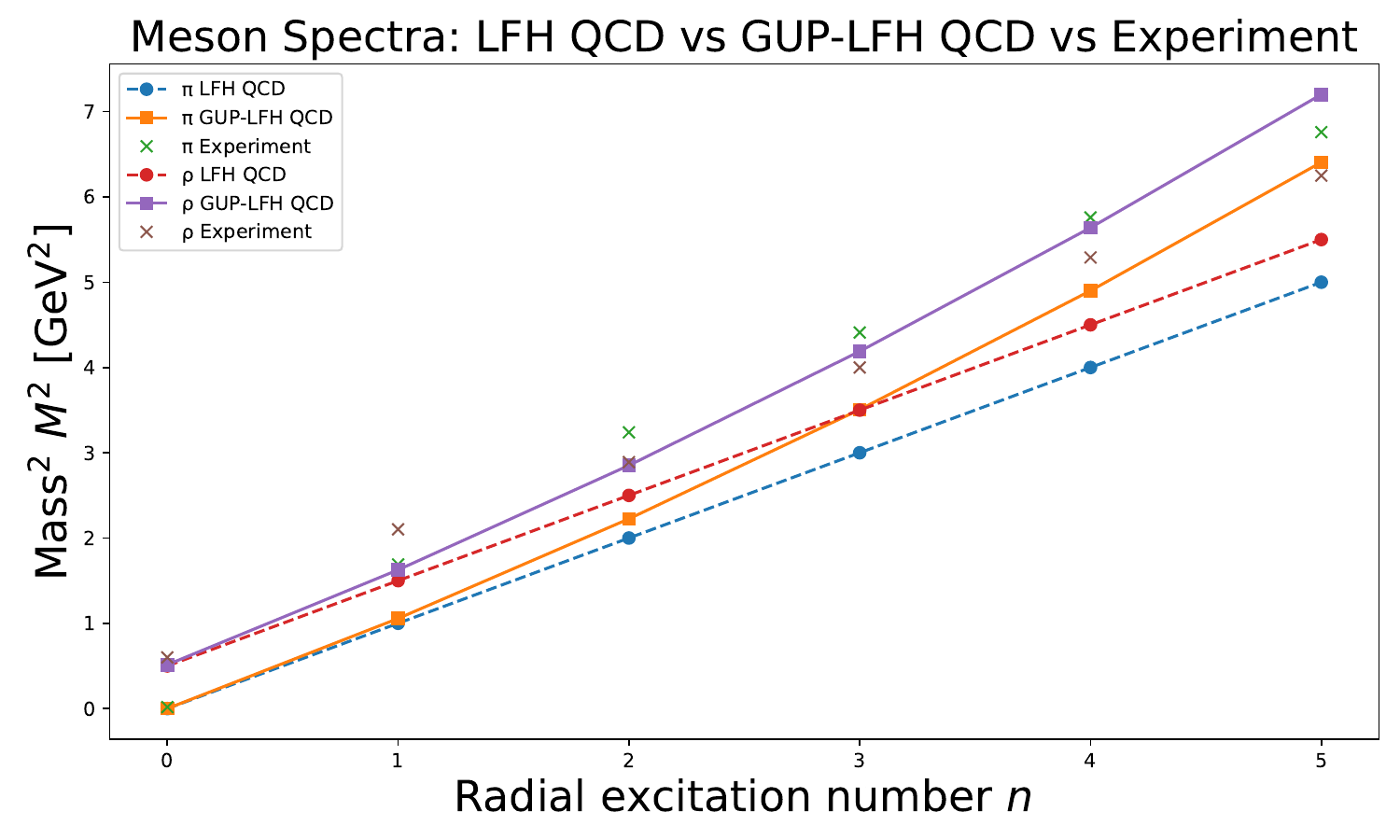}
    \caption{The meson mass spectrum ${M^{2}}$ as a function of the radial quantum number $n$, including experimental data and theoretical predictions with and without GUP corrections. The values of parameters used are $\kappa = 0.5$ and $\beta = 0.65$ GeV$^{-2}$.}
    \label{fig:2}
\end{figure}

\begin{figure}[htb]
    \centering    \includegraphics[scale=0.325]{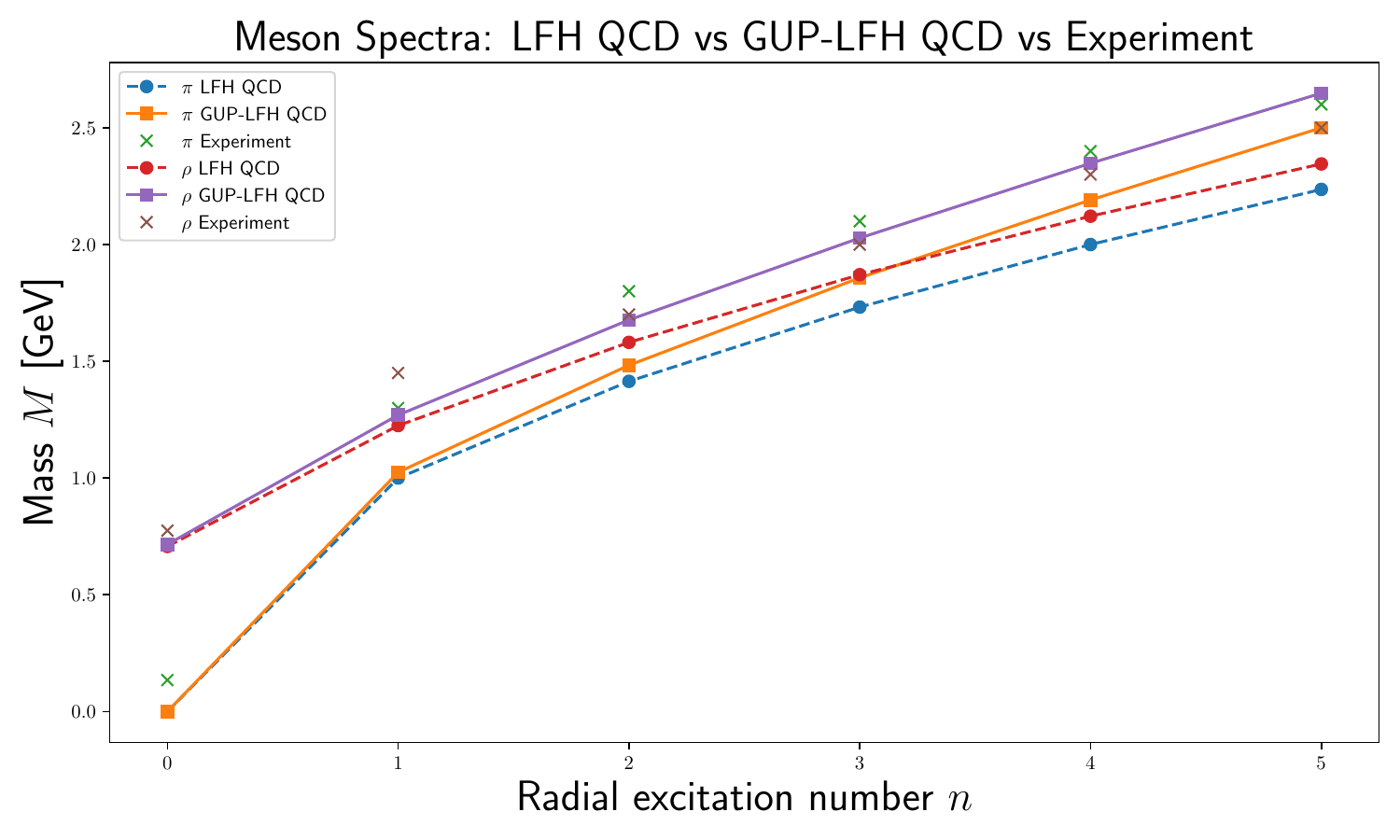}
    \caption{The meson mass spectrum $\sqrt{M^{2}}$ as a function of the radial quantum number $n$, including experimental data and theoretical predictions with and without GUP corrections. The values of parameters used are $\kappa = 0.5$ and $\beta = 0.65$ GeV$^{-2}$.}
    \label{fig:3}
\end{figure}
Next, we fit the mass spectrum—with and without GUP corrections—as a function of the radial quantum number $n$ to the experimental data for the $\pi$ and $\rho$ mesons, and deduce the value of $\beta$. All fitting plots are obtained using a confinement scale of $\kappa = 0.5~\text{GeV}$. The masses of the $\pi$ and $\rho$ mesons are listed in Table~\eqref{tab:meson_masses}, and are taken from the Particle Data Group (PDG)~\cite{PDG2022}. 
Figure~\eqref{fig:2} and Figure~\eqref{fig:3} show the plots of  $M^{2}$ and $M$ with and with GUP corrections, respectivelly, as function of the radial quantum number $n$. Both plots show that the GUP corrected- mass spectrum describes much better the experimental data. 
The value of $\beta$ that corresponds to the best fit  is $\beta = 0.65~\text{GeV}^{-2}$. Assuming that the GUP corrections are the only contributions needed to match the experimental data, we find that the length scale that corresponds to the best fit would be approximately $1.65 \times 10^{-16}~\text{m}$, which is slightly smaller than the QCD scale. This suggests that light mesons would be particularly sensitive to GUP corrections.
\section{Conclusions}\label{sec:concl5}
In this article, we have developed a novel framework for light-front holographic QCD (LFHQCD) by incorporating the Generalized Uncertainty Principle (GUP) into the QCD Lagrangian, introducing a minimal length scale associated with quantum gravitational effects. By modifying the canonical commutation relations, we derived a GUP-corrected LFHQCD equation, which manifests as a light-front Schrödinger-like equation with a higher-derivative term. This term, proportional to the GUP parameter \(\beta\), alters the ultraviolet behavior of the hadronic wave functions, leading to an enhanced mass spectrum characterized by an additional contribution $\delta M^2$.

Our phenomenological analysis demonstrates that the GUP-corrected mass spectrum significantly improves agreement with experimental data for light mesons such as the $\pi$ and $\rho$. By fitting the modified spectrum to PDG masses~\eqref{tab:meson_masses}, we determine an optimal value of $\beta = 0.65 \, \text{GeV}^{-2}$, corresponding to a minimal length scale of approximately $1.65 \times 10^{-16} \, \text{m}$, which is remarkably close to the QCD confinement scale. This suggests that light mesons may be sensitive to GUP-induced modifications. However, the fact that this minimal length is comparable to the QCD scale indicates that additional non-GUP contributions are likely involved; hence, the extracted value should be interpreted as an upper bound, i.e., $\beta < 0.65~\text{GeV}^{-2}$.

In effective GUP models, a deformation with $\beta \sim 0.65~\text{GeV}^{-2}$ does not represent fundamental quantum gravity effects directly. Instead, it mimics the formal structure of quantum gravity-motivated minimal length scenarios to effectively encode short-distance modifications of QCD dynamics. In this sense, the GUP serves as a phenomenological tool for modeling UV-regularization and hadronic structure, rather than a description of quantum gravity itself.

The success of the GUP-corrected LFHQCD framework in reproducing Regge-like trajectories of light mesons suggests that mimicking quantum gravity effects through GUP at hadronic scales can provide meaningful insights into the interplay between strong interactions and short-distance physics. By preserving confinement via the soft-wall model’s harmonic potential and incorporating UV regularization through the GUP deformation, this approach offers a promising refinement to hadron spectroscopy models.

Future work should focus on extending this framework to baryons and exotic states, incorporating additional QCD corrections such as one-gluon exchange for hyperfine splittings, and exploring alternative GUP formulations (e.g., non-quadratic forms) to better constrain $\beta$.  

This study opens a promising avenue for probing the interface between quantum gravity and hadron physics by effectively mimicking minimal length effects through GUP-inspired modifications. It encourages further exploration of how such effective models can capture aspects of short-distance physics in strongly coupled systems.

\begin{acknowledgments}
F.T. would like to acknowledge the support of the National Science Foundation under grant No. PHY-
1945471.
\end{acknowledgments}

\clearpage
\hrule
\nocite{*}

\bibliographystyle{apsrev4-2}
\bibliography{apssamp}

\end{document}